# A perspective of twisted photonic structures

Jialin Chen[1,2], Xiao Lin[1,2,*], Mingyuan Chen[3], Tony Low[4], Hongsheng Chen[1,2,*], and Siyuan Dai[3,*]

[1]*Interdisciplinary Center for Quantum Information, State Key Laboratory of Modern Optical Instrumentation, ZJU-Hangzhou Global Scientific and Technological Innovation Center, Zhejiang University, Hangzhou 310027, China.*

[2]*International Joint Innovation Center, Key Lab. of Advanced Micro/Nano Electronic Devices & Smart Systems of Zhejiang, The Electromagnetics Academy at Zhejiang University, Zhejiang University, Haining 314400, China.*

[3]*Materials Research and Education Center, Department of Mechanical Engineering, Auburn University, Auburn, Alabama 36849, USA.*

[4]*Department of Electrical and Computer Engineering, University of Minnesota, Minneapolis, Minnesota 55455, USA.*

[*]*To whom correspondence should be addressed: xiaolinzju@zju.edu.cn (X. Lin); hansomchen@zju.edu.cn (H. Chen); sdai@auburn.edu (S. Dai)*

**Moiré superlattices—twisted van der Waals (vdW) structures with small angles—are attracting increasing attention in condensed matter physics, due to important phenomena revealed therein, including unconventional superconductivity, correlated insulating states, and ferromagnetism. Moiré superlattices are typically comprised of atomic layers of vdW materials where the exotic physics arises from the quantum electronic coupling between adjacent atomic layers. Recently, moiré electronics has motivated their photonic counterparts. In addition to vdW materials, twisted photonic systems can also be comprised of metamaterials, metasurfaces, and photonic crystals, mediated by interlayer electromagnetic coupling instead. The interplay between short-ranged interlayer quantum and long-ranged electromagnetic coupling in twisted structures are expected to yield rich phenomena in nano-optics. This perspective reviews recent progress in twisted structures for nanophotonics and outlooks emerging topics, opportunities, fundamental challenges, and potential applications.**

**Introduction**

When a graphene layer is laid upon another with a small twist angle, the composite bilayer generally reveals a moiré pattern with a period much larger than the lattice constants. The twist angle controls the interlayer quantum coupling that strongly influences electronic and optical properties of the composite bilayer. Particularly, when bilayers are twisted at the so-called magic angles, moiré flat bands would emerge and lead to unconventional superconductivity and strongly correlated electronic states in twisted bilayer graphene [1-4]. Twisted bilayer graphene and other van der Waals (vdW) moiré superlattices have become fertile ground for the exploration of exotic and emergent condensed matter physics phenomena [5-22], including fractional Chern insulators [5, 6], quantum anomalous Hall effect [7-9], ferromagnetism [10, 11], moiré ferroelectricity [12, 13], orbital magnetism [14, 15], moiré excitons [16-18], and topological physics [19-21].

The recent surge in moiré electronics inspires their counterparts in photonics, namely twisted photonic structures [23-41]. The fundamental properties of twisted photonic structures are controlled by the interlayer electromagnetic coupling in addition to the quantum electronic coupling. Due to the longer wavelength of photons and their light-matter waves (polaritons) than that of electrons, the separation between adjacent components in twisted photonic structures can range from sub-nanometer to photon wavelength [42]. The interplay between electromagnetic and quantum couplings over a broad length scale underpins rich physics in twisted photonic structures. The interlayer electromagnetic coupling introduces an additional degree of freedom in designing twisted structures with novel photonic phenomena. In addition to vdW materials, twisted photonic structures can also be comprised of artificial structures, such as metasurfaces [43, 44], metamaterials [45-48], and photonic crystals [49-53]; see Fig. 1(a)-(f). Recent theoretical and experimental works have revealed a plethora of photonic and polaritonic phenomena in twisted photonic structures [23], such as topological transition of polaritonic nano-optics [43-48], graphene plasmonic photonic crystals [24], longitudinal spin of chiral plasmons [25], moiré light lines for circular dichroism [49], two-dimensional (2D) localization-delocalization transition of light [52, 53], moiré induced photonic gauge field [26], bound state in the continuum [27], and moiré chiral metamaterials [28-30]. Therefore, twisted photonic structures are a promising platform to tailor exotic light-matter interactions and can enable a range of photonic applications, including molding the flow of light at the extreme nanoscale [23, 42, 45-48], optical sensing [28-30], chiral optics [31-34], chiral plasmonics [35-37], stereochemistry and drug development for discriminating chiral molecules of different handedness [25, 28], and novel on-chip light sources [38]. As a typical example, when compared to photonic crystal-defect laser nanocavities based on a single photonic lattice, the magic-angle lasers in nanostructured moiré superlattices has been recently reported with three salient features [38], namely smaller mode volumes, higher quality factors and smaller bandwidth, which may lead to the construction of compact and reconfigurable nanolaser arrays.

**Twisted monolayers**

We begin with an introduction of plasmonics in twisted bilayer graphene [Fig. 2]. Using scattering-type scanning near-field optical microscopy (s-SNOM) and dark-field transmission electron microscopy (TEM) [Fig. 2(a)], one can visualize the periodic variation of local surface conductivity across the moiré superlattice over the domain wall (soliton) network in bilayer graphene with small twist angles [24]. These s-SNOM results originate from the modified electronic structure by interlayer coupling and atomic reconstruction that affected plasmonic responses in bilayer graphene with small twist angles. Hence, the twisted bilayer graphene acts as a lithography-free photonic crystal for propagating surface plasmon polaritons. Recently, Ref. [24] visualized the interference pattern of propagating surface plasmon polaritons over these soliton networks [Fig. 2(b)]. Ref. [24] also predicted the formation of a plasmonic bandgap in twisted bilayer graphene that can eliminate the propagation of surface plasmon polaritons in a similar rule of the bandgap in photonic crystals [42]. However, the formation of plasmonic bandgaps in twisted bilayer graphene requires strong scattering of surface plasmon polaritons from solitons that is not yet achieved in experiments. In addition, the soliton network in twisted bilayer graphene separates AB and BA stacking domains and therefore leads to rich nonlocal responses over the moiré superlattices [39-41]. For example, by incorporating nanoelectronics measurements with the s-SNOM, Ref. [39] showed that local photocurrent varies over the moiré superlattices [Fig. 2(c)] and can be altered by electrostatic gating [39].

Chiral materials are ubiquitous in nature with right- and left-handed counterparts under the mirror symmetry. However, natural chiral materials generally have weak chiral light-matter interactions, e.g., circular dichroism. Due to the intrinsic chiral nature of interlayer quantum coupling, twisted bilayer graphene and other chiral metamaterials with moiré superlattices [28-30] are promising for chiral optics [31-34] and chiral plasmons [25, 35-37]. Ref. [25] showed that the interlayer quantum coupling in twisted bilayer graphene may be utilized to build myriads of atomically thin metasurfaces. Moreover, the effective chiral surface conductivity therein determines the dispersion of chiral plasmons and leads to a $\pm\pi/2$ phase difference between the transverse-electric (TE) and transverse-magnetic (TM) components [25]. Remarkably, this unique phase relationship for chiral plasmons results in the appearance of the unconventional longitudinal spin of plasmons [25], in addition to the conventional transverse spin [54-59], as shown in Fig. 2(d). The spin-orbit and spin-valley interactions of light are ubiquitous in photonics and plasmonics, and they have led to important phenomena, including photonic and plasmonic spin Hall effects [58-64], polarization switching of interlayer excitons [65], and near-field directionality induced by spin-momentum locking [66-70]. The longitudinal spin of plasmons is attractive and is waiting to be further explored in both theory and experiment.

Surface plasmon polaritons can become nonreciprocal (without external magnetic field) if an in-plane driving electric current is applied. This leads to the so-called plasmonic Doppler effect [71-75] that has been observed in recent experiments [76, 77]. The conventional plasmonic Doppler effect originates from the relative motion between the surface plasmon polaritons and the electric current, but the induced nonreciprocity is generally weak [71]. Ref. [71] predicted that the plasmonic nonreciprocity can be greatly enhanced in graphene moiré superlattices via the quantum plasmonic Doppler effect, which is governed by the strong electron-electron interactions in moiré superlattices where electron bands are flat with extremely low Fermi velocity [71]. This quantum plasmonic Doppler effect in graphene moiré superlattices is therefore important and is promising for nonreciprocal photonic applications.

In addition to stacked bilayers, twisted monolayers can be separated where the interlayer electromagnetic coupling surpasses the quantum electronic coupling and dominates. Under this scenario, these twisted monolayers can support surface waves with exotic isofrequency contours. To facilitate the discussion, we briefly introduce the calculation of surface wave dispersion in twisted monolayers. Without loss of generality, each monolayer is considered to be uniaxial with a conductivity tensor $\bar{\bar{\sigma}}_s$. That is, we have $\bar{\bar{\sigma}}_{s,1} = \begin{bmatrix} \sigma_{x'x',1} & 0 \\ 0 & \sigma_{y'y',1} \end{bmatrix}$ at the interface between regions 1 & 2 and $\bar{\bar{\sigma}}_{s,2} = \begin{bmatrix} \sigma_{xx,2} & 0 \\ 0 & \sigma_{yy,2} \end{bmatrix}$ at the interface between regions 2 & 3, where we refer to the coordinate system aligned with the crystal axes of the top and bottom monolayers as $x'y'z'$ and $xyz$, respectively. For surface waves propagating along the direction of $\bar{k}_{\text{in-plane}}$ in the $xy$ plane, we further introduce the $x''y''z$ coordinates so that $\bar{k}_{\text{in-plane}} = \hat{x}''k_{\text{in-plane}}$. For simplicity, here we define $\phi_0 = \delta + \pi/2$ and the azimuthal angle $\phi$ ($\phi_0$) as the angle between the $\hat{x}$ direction and $\hat{y}''$ ($\hat{y}'$) direction. Through the coordinate transformation, the conductivity of the top and bottom monolayers in the $x''y''z$ coordinates can be re-written as $\bar{\bar{\sigma}}''_{s,1} = \begin{bmatrix} \sigma''_{xx,1} & \sigma''_{xy,1} \\ \sigma''_{xy,1} & \sigma''_{yy,1} \end{bmatrix}$ and $\bar{\bar{\sigma}}''_{s,2} = \begin{bmatrix} \sigma''_{xx,2} & \sigma''_{xy,2} \\ \sigma''_{xy,2} & \sigma''_{yy,2} \end{bmatrix}$, respectively, where $\sigma''_{xx,1} = \sigma_{x'x',1}\sin^2(\phi - \delta) + \sigma_{y'y',1}\cos^2(\phi - \delta)$ and $\sigma''_{xx,2} = \sigma_{xx,2}\sin^2\phi + \sigma_{yy,2}\cos^2\phi$. By enforcing the electromagnetic boundary conditions, the dispersion of surface waves can be rigorously calculated [46]. The surface waves on twisted monolayers are hybrid TM-TE modes. However, for highly squeezed surface waves, the TM wave components dominate [46]. This way, the dispersion for highly-squeezed surface waves in twisted monolayers can be written as

$$\frac{\left(1+\frac{k_{z1}/\varepsilon_1}{k_{z2}/\varepsilon_2}+\frac{k_{z1}}{\omega\varepsilon_1}\cdot\sigma''_{xx,1}\right)\cdot\left(1+\frac{k_{z3}/\varepsilon_3}{k_{z2}/\varepsilon_2}+\frac{k_{z3}}{\omega\varepsilon_3}\sigma''_{xx,2}\right)}{\left(1-\frac{k_{z1}/\varepsilon_1}{k_{z2}/\varepsilon_2}+\frac{k_{z1}}{\omega\varepsilon_1}\cdot\sigma''_{xx,1}\right)\cdot\left(1-\frac{k_{z3}/\varepsilon_3}{k_{z2}/\varepsilon_2}+\frac{k_{z3}}{\omega\varepsilon_3}\sigma''_{xx,2}\right)} = e^{+i2k_{z2}d} \quad (1)$$

where $k_{z_j} = \sqrt{\frac{\omega^2}{c^2}\varepsilon_{rj} - k^2_{\text{in-plane}}}$, and $\varepsilon_{rj}$ is the relative permittivity of region $j$ ($j = 1, 2$ or $3$).

Based on equation (1), Ref. [43] showed that when two separated black phosphorus monolayers are twisted by 90°, the dispersion of surface plasmon polaritons can be engineered by tuning the chemical potential of the top and bottom black phosphorus layers. Particularly, the inner-most isofrequency contour can change from open hyperbolas to closed ellipses by tuning the chemical potential [Fig. 3(a)]. The topological transition can also be achieved by varying the twist angle [Fig. 3(b)] between top and bottom hyperbolic metasurfaces made of monolayer graphene nanoribbons [44]. The specific twist angle at topological transition is referred to as the photonic magic angle. Both Refs. [43, 44] showed the rich dispersion engineering in twisted monolayers, which can be exploited to induce the broadband field canalization [44, 78]. The tunable topological transition via the control of chemical potential by electrostatic gating is expected to offer opportunities for active on-chip nanophotonic devices.

**Twisted slabs**

Topological transitions revealed in separated monolayers [43, 44] can also be observed in twisted slabs of $\alpha$-phase molybdenum trioxide ($\alpha$-MoO$_3$) [Fig. 1(c)]. $\alpha$-MoO$_3$ is a biaxial crystal and can support phonon polaritons with hyperbolic or elliptical dispersions in the mid-infrared region [79-86]. By using s-SNOM, four research groups [45-48] simultaneously reported polaritonic topological transitions of phonon polaritons in twisted α-MoO$_3$ slabs [Fig. 3(c)-(f)]. When the twist angle is close to the photonic magic angle, directional propagation of phonon polaritons was observed [Fig. 3(c)-(f)]. Analytical solutions for phonon polaritons in twisted α-MoO$_3$ slabs is challenging and time-consuming, due to the complexity of electromagnetic fields inside each biaxial slab and across their interface [87] involving Dyakonov surface waves [88-91]. Hence, Ref. [45, 46, 48] approximated the α-MoO$_3$ slab as an effective anisotropic monolayer whereas Ref. [47] approximated phonon polaritons in α-MoO$_3$ as pure TM guided modes. Both approximations produced theoretical results consistent with their experimental observations.

**Twisted photonic crystals**

Photonic crystals are periodic structures hosting weakly confined guided waves and can be used for electromagnetic hybrids [42]. Specifically, twisted photonic crystal [Fig. 1(d)] is a promising platform to explore the photonic analogue of moiré electronics in twisted bilayer graphene. Since twisted photonic crystals are generally quasiperiodic with a very large supercell, the computation cost is very high especially at small twist angles. This calculation difficulty impedes the systematic analysis of both photonic and electronic moiré structures. To mitigate this issue, Ref. [49] developed a high-dimensional plane wave expansion method for analyzing scattering properties of stacked photonic crystals with arbitrary twist angles [Fig. 4(a)]. This method [49] surpasses the limitations of supercell approximation and reveals strongly tunable resonant chiral behaviors in twisted photonic crystals. In particular, the concept of moiré light line has been proposed [49] and is waiting for experimental verification. In essence, the moiré light line represents the phase boundary between regions with strong and weak circular dichroism in the parameter space of twist angle and frequency [Fig. 4(b)].

The method from Ref. [49] can be used to investigate other phenomena of moiré photonics, such as photonic moiré flat bands from twisted 2D or 1D photonic crystals. The calculation of flat bands in twisted photonic structures is complex [50, 51]. Refs. [50, 51] showed the photonic band structures of twisted photonic crystals can be engineered in a similar fashion to the electronic counterpart in twisted bilayer graphene. They further discovered the photonic flat bands as a result of the interaction between in-plane and out-of-plane electromagnetic coupling [Fig. 4(c)]. In this way, the separation between photonic crystals offers a degree of freedom to tune photonic moiré bands without high pressure [50, 51]. "Magic distances" corresponding to the emergence of photonic flat bands over the whole Brillouin zone [Fig. 4(d)] has also been predicted [51].

Since twisted photonic crystals are associated with aperiodic structures and natural crystals, they can offer a feasible route to explore commensurate to incommensurate transitions [Fig. 4(e)]. Recently, reconfigurable twisted photonic crystals with controllable parameters and symmetry were created in a photorefractive crystal (e.g., strontium barium niobate (SBN): 61 crystal) by super-positioning two periodic patterns of light [52, 53]. Using these commensurable and incommensurable twisted photonic crystals, Ref. [52] observed the 2D localization-to-delocalization transition of light [Fig. 4(e)]. The observed localization of light in deterministic linear lattices is based on flat-band physics, different from previous schemes based on light diffusion in optical quasicrystals where the disorder is necessary for the onset of Anderson localization in random media. Moreover, Ref. [53] reported the formation of solitons in these moiré photonic crystal structures that smoothly evolve from fully periodic geometries to aperiodic ones, where the soliton formation is attributed to photonic flat-band physics.

## Outlook and discussion

In addition to being composed of the same material, twisted photonic structures can also be constructed with different materials [92, 93]. Intriguing routes for further exploration of twisted photonics include twisted tilted hyperbolic metamaterials [Fig. 1(e)], twisted graphene-hexagonal boron nitride heterostructures [93, 94], and twisted metasurface-photonic crystal heterostructures. Other attractive directions include twisted multilayers or twisted multi-slabs, such as twisted double bilayer graphene [95-98], twisted trilayer graphene [99], twisted multilayer metasurfaces with more complicated moiré superlattices, and even 3D photonic moiré structures with a certain periodicity in the out-of-plane direction [Fig. 1(f)].

**Innovative constituent components**

One promising direction for twisted photonics is to explore additional constituent components. For example, judiciously designed metamaterials, such as photonic hypercrystals (hyperbolic metamaterials with periodic patterns) [Fig. 5(a)-(b)], and tilted hyperbolic metamaterials with tilted optical axis [Fig. 5(c)-(d)]. Hyperbolic metamaterial [100-106] possesses hyperbolic isofrequency contours, since one component of the dielectric tensor has opposite sign to the other two. Due to this unique property, hyperbolic metamaterials support negative refraction [107, 108], superlensing effect [109], ghost polaritons [86], and diverging photonic density-of-states [110]. Particularly, hyperbolic metamaterials can enhance the recombination rate of a nearby emitter over a broad spectral range due to the high photonic density-of-states. However, this does not directly lead to a high quantum efficiency since the excited high-momentum eigenmodes are generally confined in the metamaterial. Ref. [111] overcomes this issue by building photonic hypercrystals which possess combined virtues of large broadband photonic density-of-states in hyperbolic metamaterials and strong light outcoupling in photonic crystals [Fig. 5(a)-(b)]. Correspondingly, the large enhancement of light emission was observed [111]. Moreover, Ref. [112] theoretically reported that the broadband enhancement of on-chip photon extraction can also be achieved by tilting hyperbolic metamaterials with respect to the end-facet of nanofibers [Fig. 5(c)-(d)]. Remarkably, the eigenmodes in tilted hyperbolic metamaterials are momentum-matched with the guided waves in nanofibers. This facilitates the smooth conversion between the eigenmodes in hyperbolic metamaterials and the guided waves in nanofibers. Due to these advances, photonic hypercrystals [111, 113, 114] and tilted hyperbolic metamaterials [112] can be an attractive component for further research in twisted photonic structures. Particularly, these innovative twisted hyperbolic metamaterials are expected to host unconventional far-field phenomena related to chiral optics, different from twisted bilayer α-MoO$_3$ slabs where the near-field phenomena dominate. In addition, topological transition for surface plasmon polaritons or phonon polaritons may be explored in free-space photon or other wave systems (e.g., acoustic waves [115-119]) in twisted structures comprised of corresponding components.

**Twisted multilayers or multi-slabs**

Another prospective direction of twisted photonic structures is to investigate twisted multilayers or multi-slabs [Fig. 1(f)]. For twisted multilayers, each twist angle and distance between adjacent layers govern the interlayer interactions and therefore offer more degrees of freedom to tailor light-matter interactions in twisted photonic structures. These twisted multilayer structures can be utilized to engineer different photonic properties that are otherwise not possible in twisted bilayers, in a similar fashion as their electronic counterparts (e.g., twisted trilayer graphene, twisted double bilayer graphene) [92-99].

One prototype twisted multilayer, namely a vertical superlattice with periodic structures along the vertical direction, is sketched in Fig. 1(f). These twisted multilayers allow the investigation of exotic chiral phenomena (e.g., the photonic bandgap for circularly polarized light) and various chiral responses in structures with new governing factors (periodic, general aperiodic, and quasicrystal along the vertical direction). Chiral optics [120-128] and chiral plasmonics [129-134] for discriminating chiral molecules of different handedness [135, 136] are important in stereochemistry, biological identification, information encryption, and drug development. Twisted multilayers can be a promising platform to generate giant chiral response [36] for these purposes.

In this perspective, we have reviewed the recent progress in twisted photonic structures. One enticing short-term goal for the realm of twisted photonic structures is to explore exotic optical phenomena and functionalities that do not exist in extensively-studied material platforms (including vdW materials, metasurfaces, metamaterials, and photonic crystals) by twisting these materials. The physical origin of these phenomena and functionalities can be traced to the interlayer quantum and electromagnetic coupling in twisted photonic structures. Despite the currently rapid progresses, the realm of twisted photonic structures is relatively nascent with promising directions including the exploration of nonlocality, new components for twisted photonic structures, coupling between different polaritonic modes, and twisted multilayers. On the other hand, there are still challenges both in theory and experiment that impede the further development of this realm. One is the lack of efficient methods to calculate twisted photonic structures with a large supercell, particularly when considering the quantum interlayer coupling and the in-plane nonlocal response. Others challenges include the quality and feasibility of fabricating new twisted photonic structures (e.g., with precise control of both the interlayer twist angle and the tilting angle of each constituent layer), and further optical characterization at unprecedented subnanometer scales.

Another enticing long-term goal is to reshape the radiation feature of conventional light sources (such as moving charged particles, composite magnetic and electric dipoles, and quantum emitters) by exploiting the exotic properties of twisted photonic structures. The unique interactions between light sources and twisted photonic structures may induce a plethora of attractive yet unexplored phenomena of electromagnetic radiation [Fig. 6], such as Cherenkov radiation [137-141], transition radiation [142, 143], Smith-Purcell radiation [144-147], spin-related near-field directionality [59, 67, 69, 70], and anomalous Doppler effect [60, 61], which may enable novel on-chip applications for conventional light sources.

**Acknowledgement**

The work at Zhejiang University was sponsored by the National Natural Science Foundation of China (NSFC) under Grants No. 61625502, No.11961141010, No. 61975176 and No. 62175212, the Top-Notch Young Talents Program of China, the Fundamental Research Funds for the Central Universities (2021FZZX001-19), and Zhejiang University Global Partnership Fund. The work in S.D.'s group was supported by National Science Foundation under Grant No. DMR-2005194.

**Data availability**
Data sharing is not applicable to this article as no new data were created or analyzed in this study.

**Conflict of interest**
The authors have no conflicts to disclose.

[41] N. C. H. Hesp, I. Torre, D. R. Legrain, P. Novelli, Y. Cao, S. Carr, S. Fang, P. Stepanov, D. B. Ruiz, H. H. Sheinfux, K. Watanabe, T. Taniguchi, D. K. Efetov, E. Kaxiras, P. J. Herrero, M. Polini and F. H. L. Koppens, "Observation of interband collective excitations in twisted bilayer graphene," Nature Physics **17**, 1162 (2021).
[42] J. Joannopoulos, S. Johnson, J. Winn, and R. Meade, *Photonic Crystals: Molding the Flow of Light*, (Princeton Univ. Press, Princeton, NJ, 2011).
[43] M. Renuka, X. Lin, Z. Wang, L. Shen, B. Zheng, H, Wang, and H. Sheng, "Dispersion engineering of hyperbolic plasmons in bilayer 2D materials," Optics Letters **43**, 5737-5740 (2018).
[44] G. Hu, A. Krasnok, Y. Mazor, C. Qiu, and A. Alù, "Moiré hyperbolic metasurfaces," Nano Letters **20**, 3217-3224 (2020).
[45] G. Hu, Q. Ou, G. Si, Y. Wu, J. Wu, Z. Dai, A. Krasnok, Y. Mazor, Q. Zhang, Q. Bao, C. Qiu, and A. Alù, "Topological polaritons and photonic magic angles in twisted α-$MoO_3$ bilayers," Nature **582**, 209-213 (2020).
[46] M. Chen, X. Lin, T. H. Dinh, Z. Zheng, J. Shen, Q. Ma, H. Chen, P. J. Herrero, and S. Dai, "Configurable phonon polaritons in twisted α-$MoO_3$," Nature Materials **19**, 1307-1311 (2020).
[47] Z. Zheng, F Sun, W. Huang, J. Jiang, R. Zhan, Y. Ke, H. Chen, and S. Deng, "Phonon polaritons in twisted double-layers of hyperbolic van der waals crystals," Nano Letters **20**, 5301-5308 (2020).
[48] J. Duan, N. C. Robayna, J. T. Gutierrez, G. Perez, I. Prieto, J. Sanchez, A. Nikitin, and P. Gonzalez, "Twisted nano-optics: manipulating light at the nanoscale with twisted phonon polaritonic slabs," Nano Letters **20**, 5323-5329 (2020).
[49] B. Lou, N. Zhao, M. Minkov, C. Guo, M. Orenstein, and S. Fan, "Theory for twisted bilayer photonic crystal slabs," Physical Review Letters **126**, 136101 (2021).
[50] K. Dong, T. Zhang, J. Li, Q. Wang, F. Yang, Y. Rho, D. Wang, C. Grigoropoulos, J. Wu, and J. Yao, "Flat bands in magic-angle bilayer photonic crystals at small twists," Physical Review Letters **126**, 223601 (2021).
[51] D. X. Nguyen, X. Letartre, E. Drouard, P. Viktorovitch, H. Nguyen, and H. Nguyen, "Magic configurations in moiré superlattice of bilayer photonic crystal: almost-perfect flatbands and unconventional localization," arXiv: 2104.12774v3 (2021).
[52] P. Wang, Y. Zheng, X. Chen, C. Huang, Y. V. Kartashov, L. Torner, V. V. Konotop, and F. Ye, "Localization and delocalization of light in photonic moiré lattices," Nature **577**, 42-46 (2020).
[53] Q. Fu, P. Wang, C. Huang, Y. V. Kartashov, L. Torner, V. V. Konotop, and F. Ye, "Optical soliton formation controlled by angle twisting in photonic moiré lattices," Nature Photonics **14**, 663-668 (2020).
[54] K. Y. Bliokh and F. Nori, "Transverse spin of a surface polariton," Physical Review A **85**, 061801(R) (2012).
[55] K. Y. Bliokh, A. Y. Bekshaev, and F. Nori, "Extraordinary momentum and spin in evanescent waves," Nature Communications **5**, 3300 (2014).
[56] J. S. Eismann, L. H. Nicholls, D. J. Roth, M. A. Alonso, P. Banzer, F. J. Rodríguez-Fortuño, A. V. Zayats, F. Nori, and K. Y. Bliokh, "Transverse spinning of unpolarized light," Nature Photonics **15**, 156-161 (2021).
[57] T. Mechelen and Z. Jacob, "Universal spin-momentum locking of evanescent waves," Optica **3**, 118 (2016).
[58] K. Y. Bliokh, D. Smirnova, and F. Nori, "Quantum spin Hall effect of light," Science **348**, 1448 (2015).
[59] K. Y. Bliokh, F. J. Rodríguez-Fortuño, F. Nori, and A. V. Zayats, "Spin-orbit interactions of light," Nature Photonics **9**, 796-808 (2015).
[60] X. Shi, X. Lin, I. Kaminer, F. Gao, Z. Yang, J. D. Joannopoulos, M. Soljačić, and B. Zhang, "Superlight inverse Doppler effect," Nature Physics **14**, 1001-1005 (2018).
[61] X. Lin and B. Zhang, "Normal Doppler frequency shift in negative refractive-index systems," Laser & Photonics Review **13**, 201900081 (2019).
[62] X. Zhou, X. Lin, Z. Xiao, T. Low, A. Alù, B. Zhang, and Handong Sun, "Controlling photonic spin Hall effect via exceptional points," Physical Review B **100**, 115429 (2019).
[63] S. Li, Z. Chen, L. Xie, Q. Liao, X. Zhou, Y. Chen, and X. Lin, "Weak measurements of the waist of an arbitrarily polarized beam via in-plane spin splitting," Optics Express **29**, 8777-8785 (2021).
[64] W. Kamp, F. Culchac, R. B. Capaz, and F. A. Pinheiro, "Photonic spin Hall effect in bilayer graphene moiré superlattices," Physical Review B **98**, 195431 (2018).
[65] A. Ciarrocchi, D. Unuchek, A. Avsar, K. Watanabe, T. Taniguchi, and A. Kis, "Polarization switching and electrical control of interlayer excitons in two-dimensional van der Waals heterostructures," Nature Photonics **13**, 131-136 (2019).
12

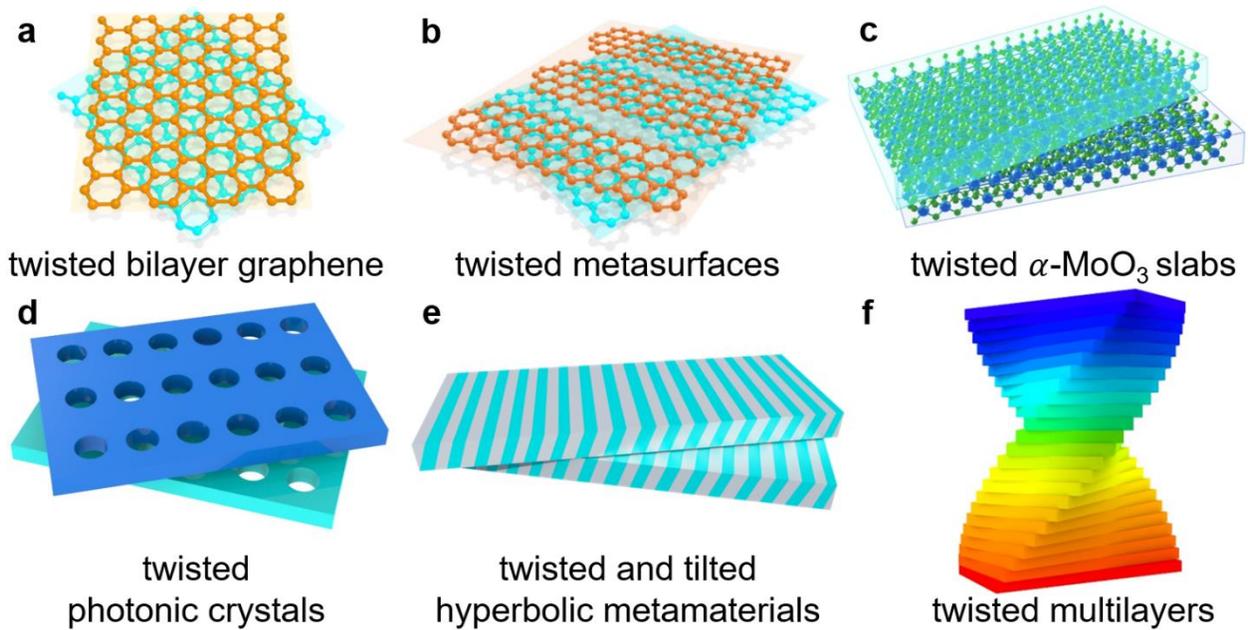

**Figure 1. Schematic of various twisted photonic structures.** (a) Twisted bilayer graphene. (b) Twisted metasurfaces. The metasurface is anisotropic, which can be constructed by, for example, monolayer black phosphorus or arrays of graphene nanoribbons. (c) Twisted α-MoO$_3$ slabs where α-MoO$_3$ is a biaxial material. (d) Twisted photonic crystals. (e) Twisted and tilted hyperbolic metamaterials. The optical axis of each hyperbolic metamaterial is tilted by a nonzero angle with respect to the vertical axis. (f) Twisted multilayers or multi-slabs.

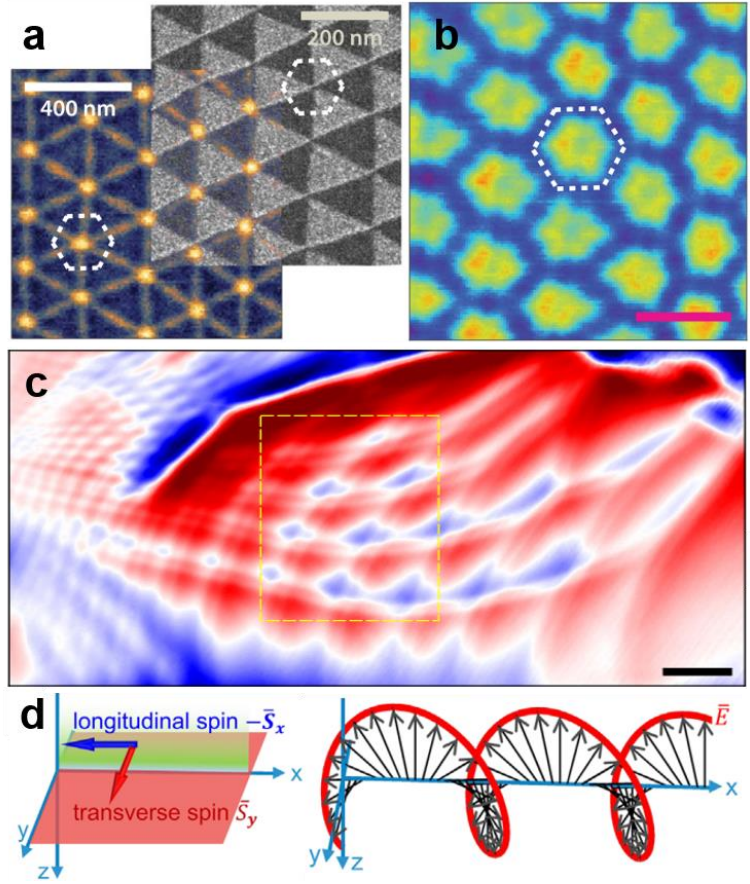

**Figure 2. Plasmonics in twisted bilayer graphene.** (a) (Left) Nano-light photonic crystal formed by a network of soliton lattices in twisted bilayer graphene [24]. The contrast is due to the enhanced local optical conductivity at solitons. (Right) Dark-field TEM image of twisted bilayer graphene, which shows the contrast between domains with different stackings, namely AB or BA stackings. The dashed hexagons represent unit cells. (b) Near-field phase image of twisted bilayer graphene at $\omega = 905$ cm$^{-1}$ with a plasmon wavelength of $\lambda_p = 282$ nm [24]. (c) Photocurrent map of minimally twisted bilayer graphene ($\theta < 0.1°$) with a carrier density of $1 \times 10^{12}$ cm$^{-2}$ and optical excitation energy of 188 meV [39]. The scalebar is 500 nm. (d) Longitudinal spin of chiral plasmons supported by twisted bilayer graphene [25]. Panels (a)-(b) are reproduced with permission from S. S. Sunku *et al.*, Science **362**, 1153-1156 (2018); Copyright 2018 American Association for the Advancement of Science. (c) is reproduced with permission from N. C. H. Hesp *et al.*, Nature Communications **12**, 1640 (2021); Copyright 2021 Springer Nature. (d) is reproduced with permission from X. Lin *et al.*, Physical Review Letters **125**, 077401 (2020); Copyright 2020 American Physical Society.

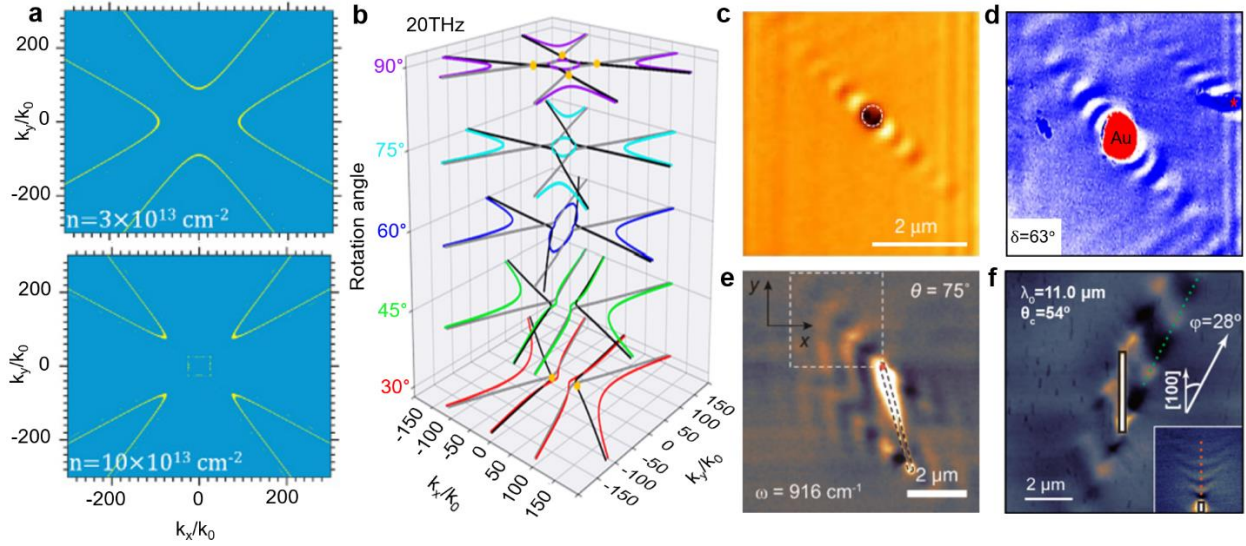

**Figure 3. Polaritonic nano-optics in twisted monolayers and twisted α-MoO₃ slabs.** (a)-(b) Engineering polariton dispersion in twisted metasurfaces [43, 44]. The metasurface is hyperbolic, made of monolayer black phosphorous in (a) or arrays of graphene nanoribbons in (b). The topological transition of isofrequency contours is controlled by tuning the chemical potential of black phosphorus (with a fixed interlayer twist angle of 90º) in (a) or by changing the interlayer twist angle in (b). The dispersion is featured with flattened isofrequency contours at the photonic magic angle, namely the specific twist angle where the topological transition occurs, as shown in (b). (c)-(f) Near-field optical images of directional phonon polaritons in twisted α-MoO₃ slabs [45-48]. Panel (a) is reproduced with permission from M. Renuka *et al.*, Optics Letters **43**, 5737-5740 (2018); Copyright 2018 OSA Publishing Group. (b) is reproduced with permission from G. Hu *et al.*, Nano Letters 20, 3217-3224 (2020); Copyright 2020 ACS Publications. (c) is reproduced with permission from G. Hu *et al.*, Nature **582**, 209-213 (2020); Copyright 2020 Springer Nature. (d) is reproduced with permission from M. Chen *et al.*, Nature Materials **19**, 1307-1311 (2020); Copyright 2020 Springer Nature. (e) is reproduced with permission from Z. Zheng *et al.*, Nano Letters **20**, 5301-5308 (2020); Copyright 2020 ACS Publications. (f) is reproduced with permission from J. Duan *et al.*, Nano Letters **20**, 5323-5329 (2020); Copyright 2020 ACS Publications.

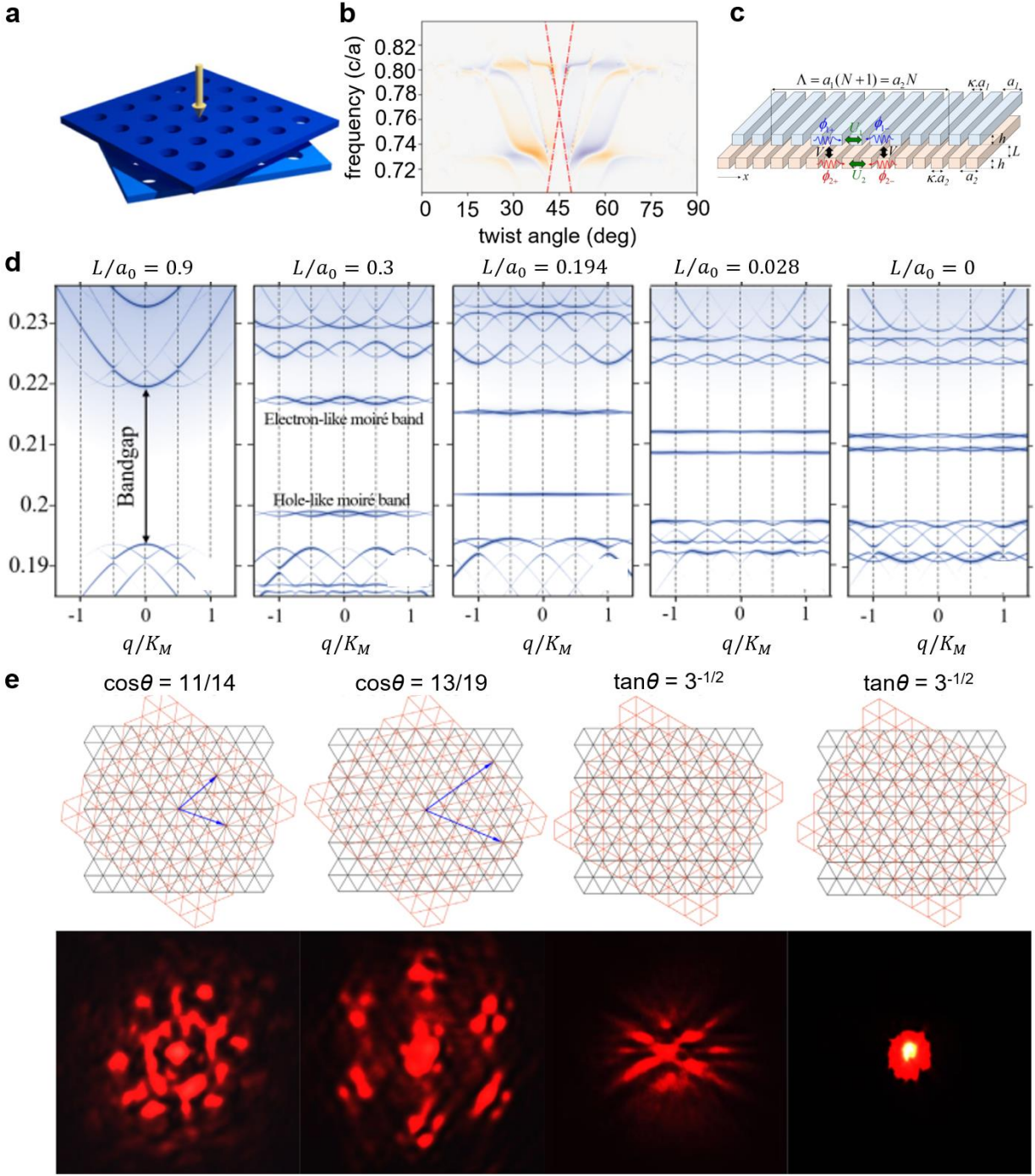

**Figure 4. Chirality, flat band, and localization of light in twisted photonic crystals.** (a) Schematic of twisted 2D photonic crystals, illuminated with normally incident light [49]. (b) Circular dichroism as a function of frequency and twist angle [49]. The chiral response of twisted photonic crystals is significantly weaker in the region between the two red dashed lines in (b), which are referred as the moiré light line. (c) Schematic of moiré

superlattice of two 1D photonic crystals (namely gratings) with periods $a_1$ and $a_2$ satisfying the condition of $a_1/a_2 = N/(N + 1)$, where $N$ is an integer number [51]. These 1D photonic crystals are separated by a subwavelength distance of $L$. (d) Simulated band structures of moiré superlattice corresponding to different $L$ values [51]. Here, $a_0 = (a_1 + a_2)/2$. Flat moiré bands would appear by tuning the separation distance $L$ in order to control the interplay between intralayer and interlayer coupling. (e) Moiré lattices created by the superposition of two rotated hexagonal lattices [52]. The top row shows the schematic discrete representation of two twisted hexagonal sublattices. The bottom row shows the measured output-intensity distributions for the transmitted signal beam at the output face of moiré lattices. The 2D localization-delocalization transition of light is experimentally observed by using commensurable and incommensurable moiré superlattices in (e). Panel (a)-(b) are reproduced with permission from B. Lou *et al.*, Physical Review Letters **126**, 136101 (2021); Copyright 2021 American Physical Society. (c)-(d) are adapted from D. X. Nguyen *et al.*, arXiv e-prints https://arxiv.org/abs/2104.12774; licensed under arXiv.org Creative Commons Attribution. (e) is reproduced with permission from P. Wang *et al.*, Nature **577**, 42-46 (2020); Copyright 2020 Springer Nature.

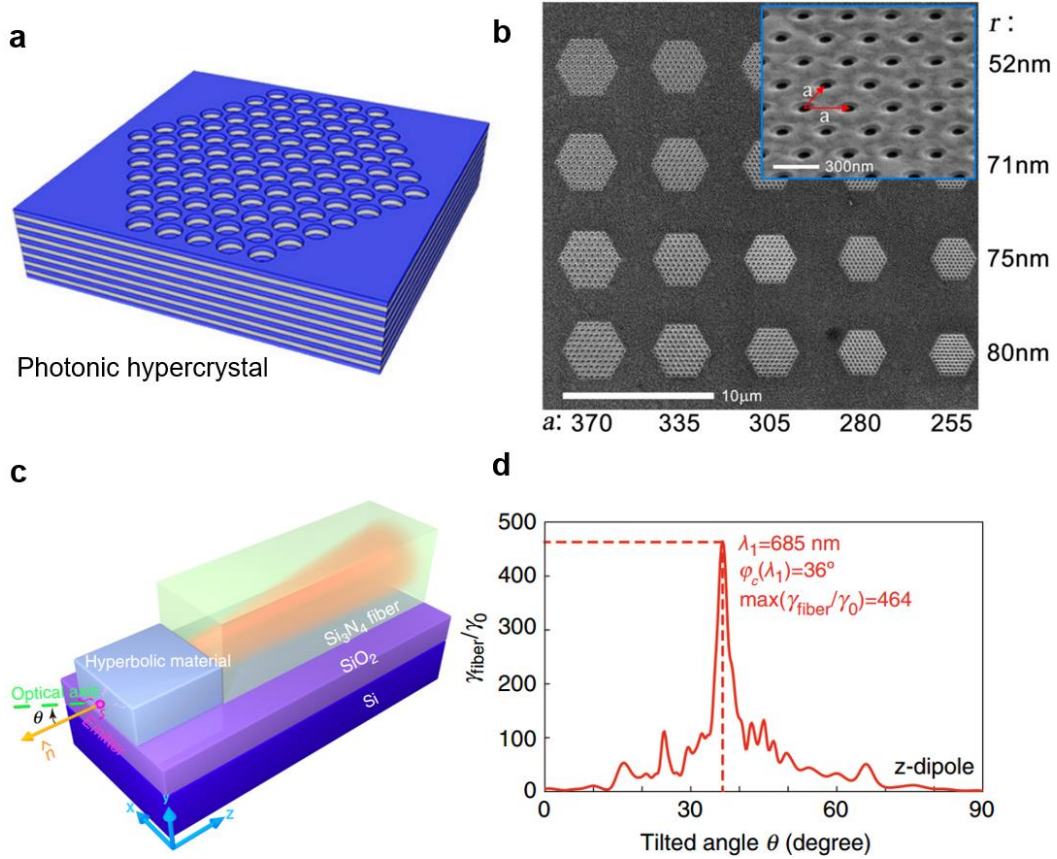

**Figure 5. New structures for twisted photonics by exploiting photonic hypercrystals and tilted hyperbolic metamaterials.** (a)-(b) Photonic hypercrystals, namely hyperbolic metamaterials with periodic patterns, such as that with hole arrays [111]. One typical photonic hypercrystal is shown in (a), with the corresponding SEM image of the fabricated sample with different pitches in (b) [111]. (c)-(d) Tilted hyperbolic metamaterial integrated with a nanofiber [112]. The optical axis of hyperbolic metamaterial forms a tilted angle $\theta$ with respect to the norm vector of interface, as shown in (c). The normalized on-chip photon extraction rate $\gamma_{\text{fiber}}/\gamma_0$ of a quantum emitter, which is positioned close to the hyperbolic metamaterial, is plotted as a function of the tilted angle in (d) [112]. Both photonic hypercrystals and tilted hyperbolic metamaterials can be used to enhance broadband spontaneous emission and light outcoupling. Panel (a)-(b) are reproduced with permission from T. Galfsky *et al.*, Proceedings of the National Academy of Sciences **114**, 5125-5129 (2017); Copyright 2017 National Academy of Sciences. (c)-(d) are reproduced with permission from L. Shen *et al.*, Applied Physics Reviews **7**, 021403 (2020); Copyright 2020 AIP Publishing LLC.

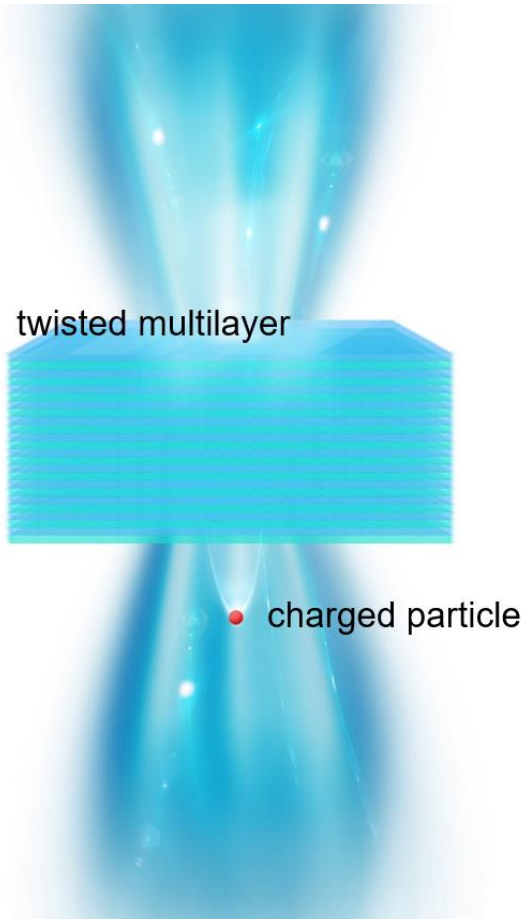

**Figure 6. Schematic of radiation induced by the interaction between various light sources and twisted photonic structures.** The light source can be, for example, moving charges, composite electric and magnetic dipoles (e.g., circularly polarized dipole, Huygens dipole and Janus dipole [59]), and quantum emitters. For conceptual illustration, this figure shows the free-electron radiation when a charged particle perpendicularly penetrates through a twisted multilayer.